\documentclass[preprint,aps,floatfix]{revtex4-1}

\usepackage{graphicx} 
\usepackage{graphics} 
\usepackage{color} 
\usepackage{subfigure} 
\usepackage{amsmath} 
\usepackage{amssymb, amsthm} 
\usepackage{txfonts} 

\begin{document}

\title{The Stochastic Gross--Pitaevskii Methodology}

%%
%%---- Authors and affiliations --------------------------------------------------------------------------
%%
\author{Stuart P. Cockburn}% \\
\author{Nick P. Proukakis}

\affiliation{Joint Quantum Centre (JQC) Durham-Newcastle,\\ 
School of Mathematics and Statistics, Newcastle University,\\ 
Newcastle upon Tyne, NE1 7RU, United Kingdom
}

\begin{abstract}
We review the \index{Gross--Pitaevskii equation! stochastic}stochastic Gross--Pitaevskii approach for \index{non-equilibrium!Bose gas}non-equilibrium finite temperature Bose gases, focussing on the formulation of Stoof; this method provides a unified description of condensed and thermal atoms, and can thus describe the physics of the critical fluctuation regime. We discuss simplifications of the full theory, which facilitate straightforward numerical implementation, and how the results of such \index{stochastic!simulation}stochastic simulations can be interpreted, including the procedure for extracting \index{coherence! phase}phase-coherent (`condensate') and \index{coherence! density}density-coherent (`\index{quasi-condensate}quasi-condensate') fractions.  The power of this methodology is demonstrated by successful {\it ab initio} modelling of several recent \index{atom chip}atom chip experiments, with the important information contained in each individual realisation highlighted by analysing \index{soliton!dark}\index{dark soliton|see{soliton, dark}}dark soliton decay within a phase-fluctuating condensate.\\~\\
Unedited version of chapter to appear in\\
Quantum Gases: Finite Temperature and Non-Equilibrium Dynamics (Vol. 1 Cold Atoms Series).\\
N.P. Proukakis, S.A. Gardiner, M.J. Davis and M.H. Szymanska, eds.\\ Imperial College Press, London (in press).\\
See http://www.icpress.co.uk/physics/p817.html
\vspace{-5cm}
\end{abstract}
\maketitle

\section{Introduction}
Many theories have been devised to study the static and dynamic 
properties of weakly interacting, ultracold, atomic Bose gases at
finite temperatures \cite{proukakis_jackson_08,blakie_bradley_08}.
An important feature of these systems is that a condensate coexists with a non-condensed component beneath the
temperature for the \index{onset!of condensation}onset of Bose--Einstein condensation (BEC):
repulsive interatomic interactions cause a depletion of atoms from the condensate, while thermal effects  additionally promote atoms from the ground state of the system. An accurate description of
partially condensed Bose gases thus requires a theory capable of describing 
condensed and non-condensed fractions within a unified framework.
Symmetry breaking finite-temperature approaches
rely on the explicit existence of a condensate mean field; such approaches are therefore useful
away from the critical region, where there is a 
well-established condensate. Nonetheless,
at temperatures close to the transition point, critical 
\index{fluctuations!thermal}fluctuations make the definition of a condensate mean field much more difficult,
an important consideration
when studying condensate growth, or low dimensional geometries, for which the temperature range over which \index{fluctuations!thermal}fluctuations are important is broader.
Fluctuations
play a key role in such cases and
their presence motivates a \index{stochastic!method}stochastic description 
of the weakly interacting Bose gas. 
While several such methods are discussed in
this book, this Chapter focuses
on the \index{non-equilibrium!quantum field theory}non-equilibrium formulation of Stoof \cite{stoof_97,stoof_chapter_98,stoof_99,stoof_bijlsma_01,duine_stoof_01}, 
whose end result is a nonlinear \index{Langevin}Langevin equation for the field representing the condensate and its \index{fluctuations!thermal}fluctuations.

\section{Methodology} 
To treat the \index{fluctuations!thermal}fluctuations inherent to the process of
Bose--Einstein condensation,
a \index{non-equilibrium!Bose gas}non-equilibrium probabilistic theory is required.
\index{Fokker--Planck}Fokker--Planck equations,  
originally introduced to describe the Brownian
motion of particles,  
achieve this by describing the time-evolution of 
an appropriate probability distribution. The mapping 
of a \index{Fokker--Planck}Fokker--Planck equation to an equivalent representation in terms of a 
\index{Langevin}Langevin equation, is often made for numerical impementation;
such equations appear frequently in diverse fields including financial market modelling \cite{bouchaud_potters_book_00}, \index{superconductor}superconductors \cite{stephens_bettencourt_02,bustingorry_cugliandolo_06},
high-energy physics \cite{bettencourt_01} and \index{turbulence}turbulence \cite{rodean_book_96}.
\newpage

\subsection{\index{Fokker--Planck}Fokker--Planck Equation for an Ultracold Bose Gas}

In Ref.~\cite{stoof_99}, Stoof derived an equation for the evolution
of the full probability distribution for the  \index{interaction(s)!weak}weakly-interacting Bose gas 
using a field-theoretic formulation of the \index{non-equilibrium!quantum field theory}non-equilibrium \index{Keldysh}Keldysh theory \cite{danielewicz_84},
within the \index{many-body!T-matrix}many-body \index{T-matrix}T-matrix approximation.
Using a \index{mean-field theory! Hartree--Fock}Hartree--Fock-like ansatz, the total probability distribution was separated
into a product of respective probability distributions for the condensate and thermal particles,
which led to two coupled equations for these `subsystems'.
Firstly, by integrating out the thermal degrees of freedom, the dynamics of the condensate distribution function, $P[\Phi^{*},\Phi; t]$,
was found to obey:
\begin{multline}
i\hbar\frac{\partial{ }}{\partial{t}}P[\Phi^{*},\Phi;t] = \\
\begin{aligned}
&-\int d{\bf r}\ \frac{\delta}{\delta\Phi({\bf r})}
\left(-\frac{\hbar^{2}}{2m}\nabla^{2}+V_{\rm ext}({\bf r})
-iR({\bf r},t)+g|\Phi({\bf r},t)|^{2}-\mu(t)\right)\Phi({\bf r}) P[\Phi^{*},\Phi;t] \\
&+\int d{\bf r}\ \frac{\delta}{\delta\Phi^{*}({\bf r})}
\left(-\frac{\hbar^{2}}{2m}\nabla^2
+V_{\rm ext}({\bf r})+iR({\bf r},t)+g|\Phi({\bf r},t)|^{2}-\mu(t)\right)\Phi^{*}({\bf r})P[\Phi^{*},\Phi;t]
\end{aligned}
 \\
-\frac{1}{2}\int d{\bf r}\ \frac{\delta^{2}}{\delta\Phi({\bf r})\delta\Phi^{*}({\bf r})} 
\hbar\Sigma^{K}({\bf r};t) P[\Phi^{*},\Phi;t],
\label{Stoof_Fokker_Planck}
\end{multline}
where, $\delta/\delta\Phi$ represents a functional derivative with
respect to the complex field $\Phi$. 
The term $iR({\bf r},t)$ describes gain or loss of particles due to 
\index{collision(s)!condensate-thermal particle exchanging}collisions which transfer atoms between the condensate and \index{thermal!cloud}thermal cloud 
given by\index{interaction(s)!between coherent and incoherent regions}
\begin{multline}
R({\bf r},t)=2\pi g^{2} 
\int\frac{d{\bf p}_{2}}{(2\pi\hbar)^{3}}
\int\frac{d{\bf p}_{3}}{(2\pi\hbar)^{3}}
\int\frac{d{\bf p}_{4}}{(2\pi\hbar)^{3}}
(2\pi\hbar)^{3}\ \delta({\bf p}_{2}-{\bf p}_{3}-{\bf p}_{4})\\
\times\delta(\varepsilon_{c}+\tilde{\varepsilon}_{2}-\tilde{\varepsilon}_{3}-\tilde{\varepsilon}_{4})
\left[ f_{2}(f_{3}+1)(f_{4}+1)-(f_{2}+1)f_{3}f_{4} \right],
\label{R_SGPE_previousprevious}
\end{multline}
where
$
\tilde{\varepsilon}_{i}=|{\bf p}_{i}|^{2}/2m+V_{\rm ext}({\bf r})+
2g\langle|\Phi({\bf r},t)|^{2}\rangle
$ 
is the \index{mean-field theory! Hartree--Fock}Hartree--Fock energy of a thermal atom 
and $f_{i}\equiv f(\tilde{\varepsilon}_{i},t)$ 
the Wigner distribution function\index{Wigner!distribution function} for thermal atoms. 

The strength of the \index{fluctuations!thermal}fluctuations is set by
the \index{Keldysh}Keldysh self-energy, defined by
\begin{multline}
\hbar\Sigma^{K}({\bf r},t)=-4\pi i g^{2} 
\int\frac{d{\bf p}_{2}}{(2\pi\hbar)^{3}}
\int\frac{d{\bf p}_{3}}{(2\pi\hbar)^{3}}
\int\frac{d{\bf p}_{4}}{(2\pi\hbar)^{3}}
(2\pi\hbar)^{3}\ \delta({\bf p}_{2}-{\bf p}_{3}-{\bf p}_{4})\\
\times\delta(\varepsilon_{\rm c}+\tilde{\varepsilon}_{2}-\tilde{\varepsilon}_{3}-\tilde{\varepsilon}_{4})
\left[ f_{2}(f_{3}+1)(f_{4}+1)+(f_{2}+1)f_{3}f_{4} \right].
\label{Keldysh}
\end{multline}
The $R$ term arises as the 
difference between the rate of scattering processes into and out of the
low-lying modes of the system, $R=\hbar(\Gamma^{\rm out}-\Gamma^{\rm in})/2$,
where $\Gamma^{\rm out}\propto f_{2}(f_{3}+1)(f_{4}+1)$
and $\Gamma^{\rm in}\propto (f_{2}+1)f_{3}f_{4}$,
while the self-energy is the {\it sum} of these rates
$\hbar\Sigma^{K}=-i\hbar(\Gamma^{\rm out}+\Gamma^{\rm in})$.
Thus, although at equilibrium scattering processes should be zero {\it on average} ($\Gamma^{\rm out} \approx \Gamma^{\rm in}$),
\index{fluctuations!thermal}fluctuations nonetheless persist,
highlighting the {\it dynamical} description of the equilibrium state.

The above \index{Fokker--Planck}Fokker--Planck equation for the condensate
is coupled to a \index{quantum! Boltzmann equation}quantum Boltzmann equation for the 
distribution function describing the \index{thermal!cloud}thermal cloud,
\begin{equation}
\frac{\partial{f}}{\partial t}+(\nabla_{\bf p}\tilde{\varepsilon}) \cdot (\nabla f)-
(\nabla\tilde{\varepsilon}) \cdot (\nabla_{\bf p} f)=
C_{12}[f]+C_{22}[f],
\label{Stoof_Boltzmann}
\end{equation}
arising from the corresponding probability distribution evolution for thermal atoms\index{interaction(s)!within incoherent region}.
Scattering processes which transfer atoms between the condensate and thermal 
cloud are described by the \index{collisional!integral}collisional integral\index{interaction(s)!between coherent and incoherent regions}
\begin{multline}
C_{12}[f]=\frac{4\pi g^{2}}{\hbar} |\Phi|^2
\int\frac{d{\bf p}_{2}}{(2\pi\hbar)^{3}}
\int\frac{d{\bf p}_{3}}{(2\pi\hbar)^{3}}
\int\frac{d{\bf p}_{4}}{(2\pi\hbar)^{3}}
(2\pi\hbar)^{3}\ \delta({\bf p}_{2}-{\bf p}_{3}-{\bf p}_{4})\\
\times\delta(\varepsilon_{\rm c}+\tilde{\varepsilon}_{2}-\tilde{\varepsilon}_{3}-\tilde{\varepsilon}_{4})
(2\pi\hbar)^{3}\ \left[\delta({\bf p}-{\bf p}_{2})-\delta({\bf p}-{\bf p}_{3})-\delta({\bf p}-{\bf p}_{4})\right]\\
\times[(f_{2}+1)f_{3}f_{4}-f_{2}(f_{3}+1)(f_{4}+1)],
\label{R_SGPE_previous}
\end{multline}
while thermal-thermal \index{collision(s)!thermal-thermal}\index{interaction(s)!within incoherent region}collisions are represented by 
\begin{multline}
C_{22}[f]=\frac{4\pi g^{2}}{\hbar} 
\int\frac{d{\bf p}_{2}}{(2\pi\hbar)^{3}}
\int\frac{d{\bf p}_{3}}{(2\pi\hbar)^{3}}
\int\frac{d{\bf p}_{4}}{(2\pi\hbar)^{3}}
(2\pi\hbar)^{3}\ \delta({\bf p}+{\bf p}_{2}-{\bf p}_{3}-{\bf p}_{4})\\
\times\delta(\tilde{\varepsilon}+\tilde{\varepsilon}_{2}-\tilde{\varepsilon}_{3}-\tilde{\varepsilon}_{4})
[(f+1)(f_{2}+1)f_{3}f_{4}-f f_{2}(f_{3}+1)(f_{4}+1)].
\label{R_SGPE}
\end{multline}
Thus, this formalism may be interpreted as a \index{stochastic!method}stochastic number-conserving generalisation 
of the \index{ZNG}ZNG kinetic theory \cite{zaremba_nikuni_99,griffin_nikuni_09},
which supplements \index{dissipative}dissipative processes affecting the 
condensate, by essential \index{fluctuations!thermal}fluctuations.

\subsection{Formulation as a \index{Langevin}Langevin Equation}

The energy $\varepsilon_{\rm c}$ which appears in the expression for the self-energy and the \index{damping}damping term $iR({\bf r},t)$ is the energy associated with 
removing an atom from the condensate, which is
given by the operator \cite{stoof_99,stoof_bijlsma_01,duine_stoof_01}
\begin{equation}
\varepsilon_{\rm c}=-\frac{\hbar^{2}}{2m}\nabla^{2}+V_{\rm ext}({\bf r})+g|\Phi({\bf r},t)|^{2}.
\label{energy_op}
\end{equation}
As discussed by Duine and Stoof in Ref.~\cite{duine_stoof_01}, 
the fact that $\varepsilon_{\rm c}$ is an operator dependent 
upon $\Phi$ leads to a complicated
\index{stochastic!equation}stochastic equation with multiplicative \index{noise!multiplicative}noise. 
To proceed, we may approximate
the 
\index{Wigner!distribution function}Wigner functions, $f({\bf r},{\bf p},t)$, 
by \index{Bose!--Einstein distribution}Bose--Einstein distributions,
henceforth denoted by $n_{\rm BE}({\bf r},{\bf p})$.
This assumption is consistent with an equilibrium \index{thermal!cloud}thermal cloud,
which we thus represent as a \index{heat bath}heat bath with a \index{chemical potential}chemical potential, $\mu$, and 
temperature, $T$. 
This process leads to a relatively simple expression 
linking the \index{damping}damping term and the \index{Keldysh}Keldysh self-energy 
\cite{stoof_99,duine_stoof_01},
\begin{equation}
-iR({\bf r},t)=\frac{1}{4}\hbar\Sigma^{K}({\bf r},t)\left[n_{\rm BE}(\varepsilon_{\rm c})+1/2\right]^{-1}.
\label{Bose_fd}
\end{equation}
A key observation is that this represents the \index{fluctuation-dissipation relation}fluctuation-dissipation relation for the system:
it describes the relationship between the magnitude of \index{fluctuations!thermal}fluctuations, 
set by $\hbar\Sigma^{K}({\bf r},t)$, and the \index{damping}damping due
to the \index{source term}source term $iR({\bf r},t)$.
This relation depends upon the equilibrium mode populations, 
set by the sum of {\it thermal} Bose--Einstein populations $n_{\rm BE}$, and
 an extra {\it quantum} contribution of half particle per mode, {\it on average};
physically, these respectively
represent stimulated and spontaneous contributions
to the scattering rate.
As high energy atoms are here assumed to be close to \index{equilibrium!thermal}thermal equilibrium, 
Eq.\ (\ref{Bose_fd}) should be valid 
in the regime of \index{linear response}linear response, applicable to perturbations which do not 
strongly affect the \index{thermal!cloud}thermal cloud.

The \index{Fokker--Planck}Fokker--Planck equation for the condensate, Eq.\ (\ref{Stoof_Fokker_Planck}), can be mapped to an equivalent representation as 
a \index{Langevin}Langevin equation (see \cite{stoof_chapter_98}), known in this context as the \index{Gross--Pitaevskii equation! stochastic}Stochastic Gross--Pitaevskii Equation (\index{Gross--Pitaevskii equation! stochastic}SGPE), which takes the form
\begin{equation}
i\hbar \frac{\partial{\Phi({\bf r},t)}}{\partial t}=\left[-\frac{\hbar^{2}}{2m}\nabla^{2}+V_{\rm ext}({\bf r})
-iR({\bf r},t)+g|\Phi({\bf r},t)|^{2}-\mu\right]\Phi({\bf r},t)
+\eta({\bf r},t).
\label{fullsGPe}
\end{equation}
This can be identified as a $T>0$ generalisation to 
the usual Gross--Pitaevskii Equation (\index{Gross--Pitaevskii equation}GPE);
it includes the scattering of particles into/out of the \index{thermal!cloud}thermal cloud\index{interaction(s)!between coherent and incoherent regions} ($-iR\Phi$),
with condensate \index{fluctuations!thermal}fluctuations modelled via the {\it dynamical} \index{noise!dynamical}noise term $\eta$. 
In order to obtain an equation that can be easily 
solved numerically, and upon noting that
$\beta(\varepsilon_{\rm c}-\mu)$ is small
at high $T$ ($\beta=1/k_{\rm B}T$), or 
close to equilibrium ($\varepsilon_{\rm c} \approx \mu$),
we Taylor expand the \index{Bose!--Einstein distribution}Bose--Einstein distribution of
Eq.~\eqref{Bose_fd} in terms of this variable.
The result of retaining the leading order term in this expansion
is to replace the \index{fluctuation-dissipation relation}fluctuation-dissipation relation of Eq.\ (\ref{Bose_fd})
with its classical counterpart, based upon
the \index{Rayleigh--Jeans}Rayleigh--Jeans distribution,
yielding \cite{stoof_99,stoof_bijlsma_01,duine_stoof_01}
\begin{equation}
-iR({\bf r},t)=\frac{\beta}{4}\hbar\Sigma^{K}({\bf r},t)\left(\varepsilon_{\rm c}-\mu\right).
\label{class_fd}
\end{equation}
Using Eq.~\eqref{energy_op} and Eq.~\eqref{class_fd} in Eq.\eqref{fullsGPe} we finally obtain
\cite{stoof_99,stoof_bijlsma_01,duine_stoof_01}
\begin{equation}
i\hbar\frac{\partial{\Phi({\bf r},t)}}{\partial t}=\bigg(1+\frac{\beta}{4}\hbar\Sigma^{K}({\bf r},t)\bigg)
\left[-\frac{\hbar^{2}}{2m}\nabla^{2}+V_{\rm ext}({\bf r})+g|\Phi({\bf r},t)|^{2}-
\mu\right]\Phi({\bf r},t)
+\eta({\bf r},t),
\label{cSGPE}
\end{equation}
with Gaussian \index{noise!Gaussian-correlated}noise \index{ensemble! correlations}ensemble correlations
	$\langle\eta^{*}({\bf r},t)\eta({\bf r'},t')\rangle=i (\hbar^2/2)\Sigma^{K}({\bf r},t) 
	\delta({\bf r}-{\bf r'})\delta(t-t')$.
This is the form of the \index{Gross--Pitaevskii equation! stochastic}SGPE  
solved to date \cite{stoof_bijlsma_01,alkhawaja_andersen_02a,proukakis_03,proukakis_schmiedmayer_06,proukakis_06b,cockburn_proukakis_09,cockburn_nistazakis_10,cockburn_negretti_11,cockburn_gallucci_11,cockburn_nistazakis_11}.
By analogy to other \index{stochastic!method}stochastic methods,
$\Phi({\bf r},t)$ should now be understood as representing a unified description of atoms within the low-energy modes of the gas, up to some energy cutoff, that are in contact with a \index{heat bath}heat bath made up of the remaining higher energy thermal atoms.

\subsection{Stochastic \index{hydrodynamic(s)!stochastic|see{stochastic, hydrodynamics}}\index{stochastic!hydrodynamics}Hydrodynamics}

The corresponding stochastic \index{stochastic!hydrodynamics}hydrodynamic theory
led to the following generalisation of the usual continuity and Josephson equations 
\cite{duine_stoof_01}, 
\begin{gather}
	\begin{split}
  \frac{\partial n_{c}({\bf r},t)}{\partial t}+\nabla\cdot\left(n_{c}({\bf r},t){\bf v}_{\rm s}({\bf r},t)\right)
  =&
  -\frac{\beta}{2}i\Sigma^{K}({\bf r},t)\left(\mu_{\rm c}({\bf r},t)-\mu\right)n_{c}({\bf r},t)
  \\
  &+ 2\sqrt{n_{c}({\bf r},t)}\xi({\bf r},t),
	\end{split}
  \\
  \hbar\frac{\partial\theta({\bf r},t)}{\partial t}-\frac{\beta}{4}i\hbar\Sigma^{K}({\bf r},t)
	\frac{\hbar^2\nabla\cdot\left(n_{c}({\bf r},t)\nabla\theta({\bf r},t)\right)}{2m n_{c}({\bf r},t)}
  =
	\mu-\mu_{\rm c}({\bf r},t)+\frac{\nu({\bf r},t)}{\sqrt{n_{c}({\bf r},t)}},
\end{gather}
respectively, where
$n_c$ and $\theta$ are the condensate density and phase,
$
	  \mu_{\rm c}({\bf r},t)=-(\hbar^2\nabla^{2}\sqrt{n_{c}({\bf r},t)}/2m\sqrt{n_{c}({\bf r},t)})
		+V_{\rm ext}({\bf r},t)+gn_{c}({\bf r},t)+(m/2){\bf v}^{2}_{s}({\bf r},t)
$
and the velocity is given by the gradient of the phase ${\bf v}_{\rm s}({\bf r},t)
={\hbar}\nabla\theta({\bf r},t)/m$. The \index{noise}noise terms are now given by
$\langle\nu({\bf r},t)\nu({\bf r'},t')\rangle=i({\hbar^2}/{4})\hbar\Sigma^{K}({\bf r},t) 
\delta({\bf r}-{\bf r'})\delta(t-t')$
and $\langle\xi({\bf r},t)\xi({\bf r'},t')\rangle=i({\Sigma^{K}({\bf r},t)}/{4})
\delta({\bf r}-{\bf r'})\delta(t-t')$.
This approach is easily amenable to analytic variational methods \cite{duine_stoof_01,duine_leurs_04}.

\subsection{Simple Numerical Implementation}

The numerical solution of the \index{Gross--Pitaevskii equation! stochastic}SGPE, Eq.~\eqref{cSGPE},
is not much more complicated than the usual \index{Gross--Pitaevskii equation}GPE: 
following Bijlsma
and Stoof \cite{stoof_bijlsma_01,werner_drummond_97},
we seek to propagate the system from a time $t_{m}$ by a time step $\Delta t$, 
via
\begin{equation}
		\Phi({\bf r},t_{m}+\Delta t)=
\exp\left(-i[1+\beta\hbar\Sigma^{K}({\bf r})/4][\varepsilon_{\rm c}-\mu]\Delta t/\hbar\right) 
\left[\Phi({\bf r},t_{m})
		-\left(\frac{i}{\hbar}\right)\xi_{m}({\bf r})
		\right]
	\label{int_fac}
\end{equation}
where we have defined the noisy field at the $m^{\rm th}$ time step as
$
\xi_{m}({\bf r})\equiv
\exp\{-i[1+\beta\hbar\Sigma^{K}({\bf r})/4](\varepsilon_{\rm c}-\mu)t_{m}/\hbar\}
\left\{ \int_{t_{m}}^{t_{m}+\Delta t}
dt'\ \exp\{i[1+\beta\hbar\Sigma^{K}({\bf r})/4][\varepsilon_{\rm c}-\mu] t'/\hbar\} \right\} \eta{({\bf r},t')}$.
This has correlations given by
$
\langle\xi_{m}^{*}({\bf r})\xi_{n}({\bf r'})\rangle=i(\hbar^2/2)\Sigma^{K}({\bf r}) 
\delta({\bf r}-{\bf r'})\delta_{m n}\Delta t+\mathcal{O}(\Delta t^{2}).
$
Making use of Cayley's form for the exponential of Eq.~\eqref{int_fac}, the problem is then reduced 
to the solution of
\begin{multline}
		\left\{1+{i\left[1+\beta\hbar\Sigma^{K}({\bf r})/4\right]}
		\left(\varepsilon_{{\rm c},m+1/2}-\mu\right)\Delta t/{2\hbar}\right\}\Phi({\bf r},t_{m}+\Delta t) \\
   	=\left\{1-{i\left[1+\beta\hbar\Sigma^{K}({\bf r})/4\right]}\left(\varepsilon_{{\rm c},m+1/2}-\mu\right)\Delta t/{2\hbar}\right\}
		\left[\Phi({\bf r},t_{m})-\left(\frac{i}{\hbar}\right)\xi_{m}({\bf r})\right]
	\label{SGPE_sol}
\end{multline}
where $\varepsilon_{{\rm c},m+1/2}=\left[\varepsilon_{\rm c}(t_{m})+\varepsilon_{\rm c}(t_{m+1})\right]/2$ is the
operator of Eq.~\eqref{energy_op} evaluated at the mid-point of the time step \cite{werner_drummond_97}; the remaining step is to spatially discretise Eq.~\eqref{SGPE_sol}, which can be achieved using standard methods.

A typical simulation proceeds as follows: 
(i) Choose the necessary input parameters: $\mu$, $T$, atomic species, trapping potential;
(ii) Create an \index{ensemble! of realisations}ensemble of realisations, each corresponding to a unique
set of \index{noise!realisation}noise realisations at each time step; 
(iii) Propagate this set of realisations to equilibrium, i.e.\ when
observables (e.g. condensate number) become constant {\it on average};
equilibrium observables may be extracted by constructing \index{correlation functions}correlation functions
from the set of \index{noise!realisation}noise realisations, e.g. the density is given by
$\langle\Phi^{*}\Phi\rangle\equiv\sum_{i=1}^{N}\Phi_{i}^{*}\Phi_{i}/N$
where $i$ denotes a particular \index{noise!realisation}noise realisation; 
(iv) Once at equilibrium, dynamical perturbations may also be studied, 
e.g.\ topological excitations (Section\ \ref{cockburn_application2}), or collective modes.

\subsection{Interpretation: Single Runs and Extracting \index{coherence}Coherence Properties}

\subsubsection{Single Versus Averaged Runs}

By construction,
physical properties are meant to be calculated by averaging\index{stochastic!averaging} over
different realisations of the \index{stochastic!field}stochastic field $\Phi$. Nonetheless
important information may also be extracted from {\it single} numerical runs;
in this sense, the \index{Gross--Pitaevskii equation! stochastic}SGPE 
offers a strong analogy to experimental methods,
in which data is obtained through repeated measurements from several 
independent experimental realisations. The \index{Gross--Pitaevskii equation! stochastic}SGPE was first applied in this way 
to demonstrate that important details of the growth and collapse
dynamics of $^{7}$Li condensates \cite{gerton_strekalov_00} contained within {\it single} numerical realisations were lost by averaging over many runs \cite{duine_stoof_01},
suggesting that single \index{stochastic!realisation}stochastic numerical realisations are analogous to independent experimental realisations\index{stochastic!trajectory interpretation}.
The role of information extracted from single runs has been strengthened by
further analysis, including
spontaneous \index{vortex!spontaneous formation}vortex formation via the \index{Kibble--Zurek}Kibble--Zurek mechanism 
\cite{weiler_neely_08},
fluctuating \index{soliton!dark}soliton dynamics (Section \ref{cockburn_application2})
 and  
{\it in situ} \index{density!fluctuations|see{fluctuations, density}}\index{fluctuations!density}density \index{fluctuations!thermal}fluctuations in \index{atom chip}atom chip experiments (Section \ref{cockburn_application1}).

\subsubsection{{\it A Posteriori} Condensate/\index{quasi-condensate}Quasi-Condensate Extraction}
\begin{figure}[t!]
  \begin{center} 
		\includegraphics[angle=0,scale=0.23,clip]{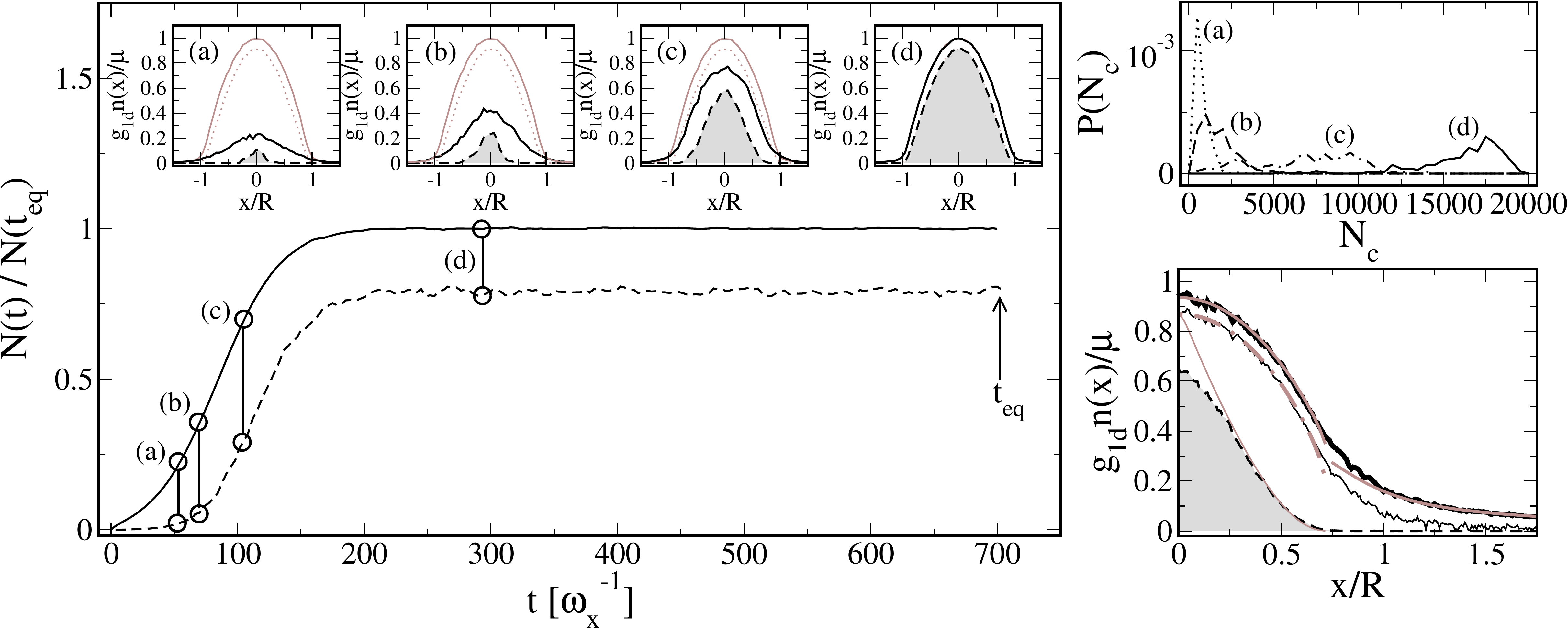} %&
    \caption[]{Left Plot: Growth of total (solid) and condensate number (dashed) during equilibration: insets show corresponding profiles at indicated times, with equilibrium results (at time $t_{\rm eq}$) shown by the thin solid and dotted lines. Right (Top): Condensate number distribution during growth at indicated times. (Bottom): Higher temperature data showing  \index{Gross--Pitaevskii equation! stochastic}SGPE total (thick black noisy line) and \index{quasi-condensate}\index{quasi-condensate!relation to Penrose--Onsager mode}quasi-condensate (thin black noisy) density profiles versus \index{modified Popov}modified Popov total (thick grey lines) and \index{quasi-condensate}quasi-condensate densities (dot-dashed grey). 
The \index{Gross--Pitaevskii equation! stochastic}SGPE Penrose--Onsager condensate\index{Penrose--Onsager!condensate mode}\index{Penrose--Onsager!condensate mode! relation to quasi-condensate}\index{Penrose--Onsager!condensate mode! extraction in c-field methods} (shaded region) is compared to the result of Eq.\ (\ref{eq:qcPO}) (thin grey line).
Figure adapted with permission from S.P. Cockburn {\em et al.}, 
{\em Comparison between microscopic methods for finite-temperature Bose gases},
Phys. Rev. A, {\bf 83}, 043619 (2011) \cite{cockburn_negretti_11}.
Copyright (2011) by the American Physical Society.
}
\label{fig:qc_PO}
\end{center}	
\end{figure}
The noisy wavefunction of the \index{Gross--Pitaevskii equation! stochastic}SGPE represents {\it both coherent} and \index{incoherent! field}{\it incoherent}
atoms within low-energy modes\index{low-lying modes} in a unified manner, and
a further
statistical analysis is required to identify the coherent components. 
The density coherent {\it \index{quasi-condensate}quasi-condensate} may be identified via 
$n_{\rm qc}(x)=\sqrt{2-g^{(2)}(x)}\,n(x)
$
\cite{prokovev_ruebenacker_01,proukakis_06b,bisset_davis_09}, 
where the second order correlation function
$g^{(2)}(x)=\langle|\Phi(x)|^4\rangle/\langle|\Phi(x)|^2\rangle^{2}$.
The additionally phase coherent fraction is associated to the
Penrose--Onsager {\it condensate} mode\index{Penrose--Onsager!condensate mode}\index{Penrose--Onsager!condensate mode! extraction in c-field methods}\index{quasi-condensate!relation to Penrose--Onsager mode} \cite{penrose_onsager_56}, identified as the eigenmode corresponding to the largest eigenvalue of the system
one-body \index{density!matrix}density matrix  
$\rho(x,x')\approx\langle\Phi^{*}(x)\Phi(x')\rangle$
\cite{blakie_bradley_08,cockburn_negretti_11,wright_proukakis_11};
this may be numerically obtained %here 
by diagonalising the \index{density!matrix}density matrix --- see Fig.~\ref{fig:qc_PO}
for an analysis of a $T>0$ 1d Bose gas based on this 
prescription.
Motivated by Refs. \cite{alkhawaja_andersen_02a,alkhawaja_andersen_02b},
the Penrose--Onsager condensate\index{Penrose--Onsager!condensate mode} of the \index{Gross--Pitaevskii equation! stochastic}SGPE is found to be well
matched within a trapped system by the definition\index{Penrose--Onsager!condensate mode! relation to quasi-condensate} \cite{cockburn_negretti_11}
\begin{equation}
n'_{\rm c}(x)=g^{(1)}(0,x)\sqrt{2-g^{(2)}(x)}\,n(x),
\label{eq:qcPO}
\end{equation}
where $g^{(1)}(0,x)=\rho(0,x)/\sqrt{n(0)n(x)}$ is the first order normalised correlation function.
The dependence upon $g^{(1)}(0,x)$ illustrates clearly the additional \index{coherence! phase}{\it phase 
coherence} of the \index{Penrose--Onsager!condensate mode}Penrose--Onsager condensate, relative to the 
\index{quasi-condensate}quasi-condensate\index{Penrose--Onsager!condensate mode! relation to quasi-condensate}\index{quasi-condensate!relation to Penrose--Onsager mode} in which only {\it \index{fluctuations!density}density \index{fluctuations!thermal}fluctuations}  are suppressed. 
Eq.~\eqref{eq:qcPO} provides an alternative means of extracting the 
phase coherent fraction of a trapped gas from \index{Gross--Pitaevskii equation! stochastic}SGPE simulations,
which accurately captures the condensate edge (but breaks down
at very small distances from the trap centre); this method is also ideal for distinguishing between `condensate' and `\index{quasi-condensate}quasi-condensate' in \index{atom chip}atom chip experiments \cite{cockburn_negretti_11}.

\subsubsection{Comparison to the \index{modified Popov}Modified Popov Method}

As an independent validation of the above interpretation, Fig.~\ref{fig:qc_PO} compares the \index{Gross--Pitaevskii equation! stochastic}SGPE result to the \index{modified Popov}\index{Popov!modified|see{modified Popov}}modified Popov theory of 
Stoof and co-workers \cite{andersen_alkhawaja_02,alkhawaja_andersen_02a,alkhawaja_andersen_02b,stoof_dickerscheid_book_09},
which accounts for contributions 
of \index{fluctuations!phase}phase \index{fluctuations!thermal}fluctuations to {\it all} orders. Densities here are obtained within the local density approximation, via
\begin{equation}
    n(x) = n_{\rm qc}(x) + 
    \frac{1}{V}\sum_{{k}}
    \left\{
    \left[n_{\rm BE}(\epsilon_{{k}})+\frac{1}{2}\right]\frac{\varepsilon_{{k}}}{\epsilon_{{k}}}
    -\frac{1}{2}
    +\frac{ g_{\rm 1d} n_{\rm qc}(x)}{2\varepsilon_{{k}} + 2 \mu}
    \right\},
\end{equation}
with 
$ 
n_{\rm qc}(x)=\left[\mu-V_{\rm ext}(x)-2g_{\rm 1d}n'(x)\right]/g_{\rm 1d}, 
$ 
where $\epsilon_{{k}} = [\varepsilon_{k}^{2} + 2g_{\rm 1d}n_{\rm qc}\varepsilon_{k}]^{1/2}$
is the \index{Bogoliubov!spectrum}Bogoliubov dispersion relation,
$\varepsilon_{k} = \hbar^2 k^2/2m$, and $V$ is the system volume.
Building on excellent agreement between the \index{Gross--Pitaevskii equation! stochastic}SGPE and \index{modified Popov}modified Popov theory
\cite{alkhawaja_andersen_02a}, and the ideas introduced in \cite{alkhawaja_andersen_02b}, the present comparison corroborates the above means of extracting the  `true' and quasi-\index{condensate!fraction}condensate fractions of the gas from \index{Gross--Pitaevskii equation! stochastic}SGPE data.
Thus, Eq.~\eqref{eq:qcPO} is also expected to be a useful 
tool for analysing {\it experimental} density profiles.
Further comparison between the \index{Gross--Pitaevskii equation! stochastic}SGPE and other \index{one-dimensional}one-dimensional Bose gases theories may be found in \cite{cockburn_negretti_11}.

\section{Validity Issues}

\subsection{Validity Domain}
Two main assumptions underlie Eq.~\eqref{cSGPE}: Firstly, high energy thermal atoms within the system are treated as being at equilibrium
(with their mean field contribution to $\Phi$ currently neglected).
\index{thermal!cloud}Thermal cloud dynamics --- crucial when the thermal cloud is strongly perturbed --- can be included by evolving the distribution functions via the \index{quantum! Boltzmann equation}quantum Boltzmann equation [Eq.\ (\ref{Stoof_Boltzmann})] \cite{stoof_99}.

Second is the so-called `classical approximation':
This terminology stems from 
the fact that the classical \index{Rayleigh--Jeans}Rayleigh--Jeans distribution arises as the leading order 
term in a small $\beta(\varepsilon_{\rm c}-\mu)$ expansion of the \index{Bose!--Einstein distribution}Bose--Einstein distribution.
While it does not constitute an essential ingredient of the theory, 
this approximation is very useful for {\em numerical} purposes, as it simplifies
the scattering term $R({\bf r},t)$ to the form of Eq.~\eqref{class_fd}, thus leading to the \index{Gross--Pitaevskii equation! stochastic}SGPE of Eq.\ (\ref{cSGPE}) that is numerically solved.
Although this is a well justified approximation for highly occupied (thus low energy) modes,
it does lead to an `ultraviolet catastrophe'\index{ultraviolet!catastrophe}, which manifests through a dependence 
of physical observables upon the energy \index{cutoff}cutoff \cite{stoof_bijlsma_01,cockburn_proukakis_09};
this problem is far more pronounced in spatial dimensions greater 
than one, due to the form of the density of states, and a possible solution 
is the introduction of divergence-cancelling counter-terms \cite{parisi_book_88}.
The \index{stochastic!field}stochastic field $\Phi$
thus represents not just the phase-coherent part (condensate), but all atoms within modes up to
an energy \index{cutoff}cutoff 
(in our simulations this is typically 
set by the spatial discretisation --- see also Section\ \ref{cockburn_application1}). 

\subsection{Relevance to Other Theories}
The \index{Gross--Pitaevskii equation! stochastic}SGPE is related to a number of theories discussed in this book:
The closest link arises to the simple growth \index{Gross--Pitaevskii equation! stochastic projected}SPGPE 
\cite{gardiner_anglin_02,gardiner_davis_03,blakie_bradley_08},
which is very similar in nature, despite their rather distinct derivations.  Various numerical differences arise in practice --- most notably the use (or not) of a \index{projector}projector to separate low and high-lying modes.  While this may be fundamentally important, the potential benefits from its use for dynamical predictions when low-lying modes are coupled to a {\it static} \index{thermal!cloud}thermal cloud are not universally accepted --- see Appendix C2 of Ref.\ \cite{proukakis_jackson_08} for a more detailed discussion of the links between these two approaches.

The \index{Gross--Pitaevskii equation! stochastic}SGPE is a grand \index{canonical}canonical theory, as the exchange of particles and energy between 
the low-energy modes and \index{heat bath}heat bath is allowed.
If, upon reaching equilibrium, $\hbar\Sigma^{K}$ is set to zero in both
 \index{damping}damping and \index{noise}noise terms (i.e.\ $iR = \eta = 0$ in Eq.~\eqref{cSGPE}), then
it reduces to a \index{multimode!field}multimode, finite temperature time-dependent \index{Gross--Pitaevskii equation}GPE
with a \index{stochastic!sampling}stochastically-sampled initial state. Such an approach,
first applied by us to study 
\index{quasi-condensate}quasi-condensate
growth on an \index{atom chip}atom chip \cite{proukakis_schmiedmayer_06}, and subsequently used for finite 
temperature \index{vortex!finite temperature dynamics}vortex dynamics \cite{rooney_bradley_10},
is similar in spirit (but not implementation \cite{proukakis_jackson_08}) to the \index{stochastic!sampling}stochastic sampling of the Wigner distribution function\index{Wigner!distribution function}
evolved in \index{truncated Wigner}truncated Wigner simulations \cite{steel_olsen_98,sinatra_lobo_02,norrie_thesis_05}.

Moreover, \index{classical field}classical field methods for Bose gases \cite{levich_yakhot_78,svistunov_91,kagan_svistunov_97,marshall_new_99,goral_gajda_01a,davis_morgan_01}, based upon the observation that the \index{Gross--Pitaevskii equation}GPE accurately describes the dynamics of all highly occupied modes,
typically start with a suitably random, \index{multimode!initial condition}multimode initial condition; this
evolves to a classical equilibrium, sampling a 
\index{microcanonical}microcanonical \index{phase-space}phase-space under \index{ergodic}ergodic \index{Gross--Pitaevskii equation} GPE evolution \cite{marshall_new_99,goral_gajda_01a,davis_morgan_01,connaughton_josserand_05}.
Since $\mu$ and $T$ are input parameters for the \index{Gross--Pitaevskii equation! stochastic}SGPE, 
the latter approach may therefore be viewed as a more controlled way of generating a
finite temperature initial state for \index{classical field}classical field simulations \cite{cockburn_negretti_11}.
However, one way in which the \index{Gross--Pitaevskii equation! stochastic}SGPE contrasts to `conventional' \index{classical field}classical field theories is through the generation of an \index{ensemble! canonical}{\it ensemble of
independent realisations}, with physical observables (such as 
\index{correlation functions}correlation functions) generated by
\index{ensemble! average}ensemble averaging over many \index{noise!realisation}noise realisations 
(although see, for example, also \cite{sinatra_lobo_01}).
In \index{classical field}classical field theories, such observables are instead typically generated by sampling
the system at many different times, chosen so as to be sufficiently far apart.

By construction, the \index{Gross--Pitaevskii equation! stochastic}SGPE incorporates \index{fluctuations!thermal}fluctuations into the condensate mean-field \index{stochastic!method}stochastically, whereas these can at most be {\it a posteriori} included in theories based on \index{symmetry-breaking}symmetry-breaking. 
Although we have focused here on numerical realisations with a static \index{thermal!cloud}thermal cloud,
the \index{Gross--Pitaevskii equation! stochastic}SGPE is in general intended to be coupled 
to a \index{quantum! Boltzmann equation}quantum Boltzmann equation for the \index{thermal!cloud}thermal cloud; in this sense the full 
theory of Stoof may be considered as an (explicitly $U(1)$-symmetry-preserving) generalisation of the \index{ZNG}ZNG method 
\cite{zaremba_nikuni_99,griffin_nikuni_09}.
The main importance of including \index{fluctuations!thermal}fluctuations is for describing dynamical processes in the region of critical \index{fluctuations!thermal}fluctuations (enhanced in \index{low-dimensional}low-dimensional systems), or for accounting for experimental shot-to-shot variations, 
whereas a \index{ZNG}ZNG-type approach could only account for {\it averaged} properties, albeit doing so very accurately.
Finally, setting $\eta \rightarrow 0$ but maintaining the \index{dissipative}dissipative contribution to Eq.\ (\ref{cSGPE}) leads to the commonly used Dissipative GPE  (\index{Gross--Pitaevskii equation! dissipative}DGPE), with an {\it ab initio} expression for the \index{damping}damping rate, as opposed to \index{phenomenological}phenomenological input.

\section{Applications}
\subsection{Comparison to Quasi-1d Bose Gas Experiments \label{cockburn_application1}}
\begin{figure}[htbp]
  \begin{center}
    \includegraphics[angle=0,scale=0.30,clip]{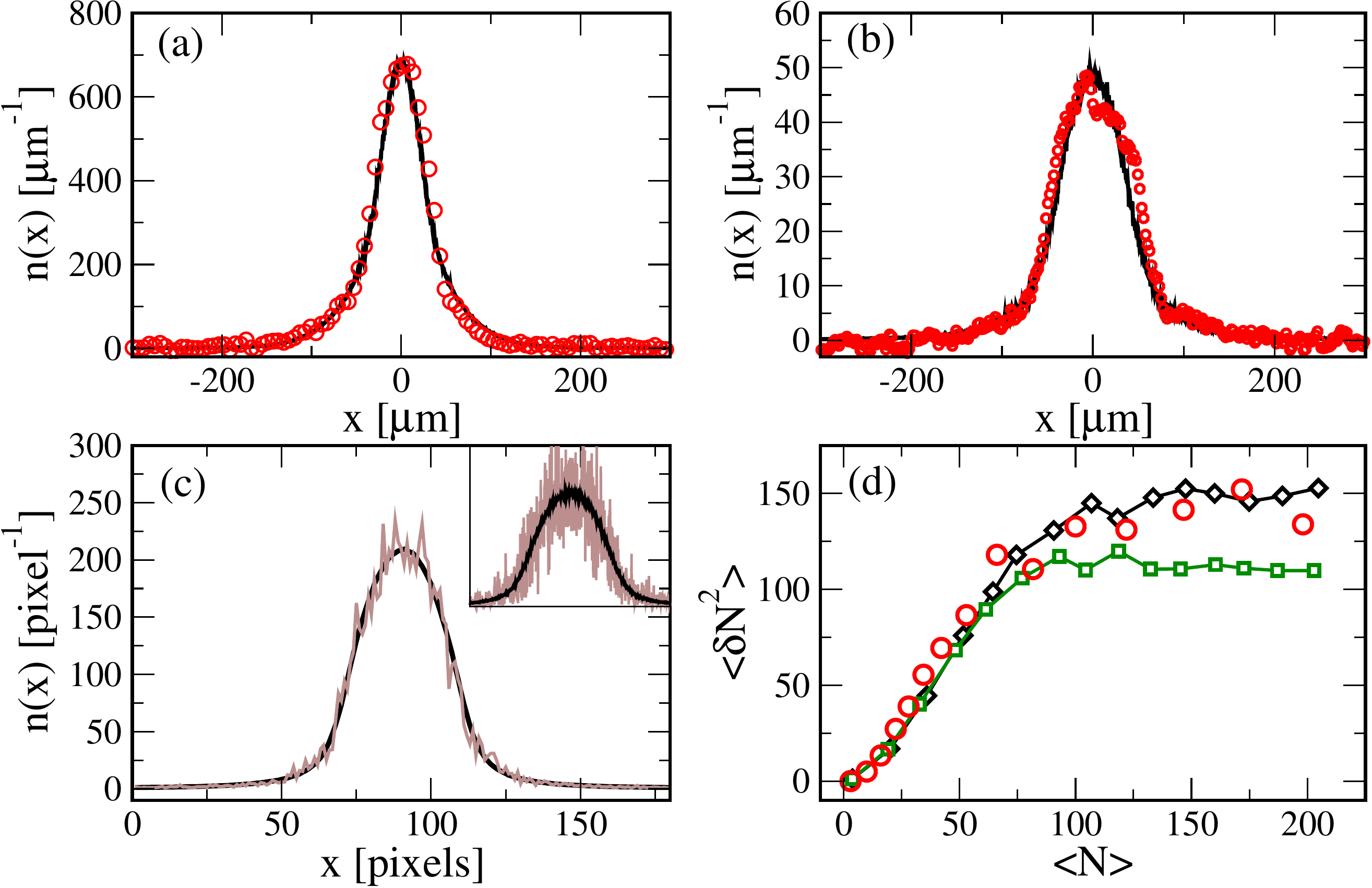}
    \caption[]{Quasi-1d \index{Gross--Pitaevskii equation! stochastic}SGPE model (solid black line) versus experimental density profiles (circles) from (a) Trebbia {\it et al.} \cite{trebbia_esteve_06} and (b) van Amerongen {\it et al.} \cite{vanamerongen_vanes_08}. (c) Main Plot: Single-run density, binned to experimental CCD camera resolution for the experiment of Armijo {\it et al.} \cite{armijo_jacqmin_10} (grey noisy) vs. average over 1000 realisations (black), with corresponding {\it raw} \index{Gross--Pitaevskii equation! stochastic}SGPE single-run data shown 
in the inset. (d) Density flutuations per binned average number predicted by the quasi-1d (black diamonds) and 1d (squares) \index{Gross--Pitaevskii equation! stochastic}SGPE [based on the binning procedure illustrated in (c)] versus experimental data of Armijo {\it et al.} (circles).
Figure adapted with permission from
S.P. Cockburn {\em et al.}, 
{\em Quantitative study of quasi-one-dimensional Bose gas experiments via the \index{Gross--Pitaevskii equation! stochastic}stochastic Gross--Pitaevskii equation}, Phys. Rev. A {\bf 84}, 023613 (2011) \cite{cockburn_gallucci_11}.
Copyright (2011) by the American Physical Society.
}
	\label{fig:q1d}
  \end{center}	
\end{figure}

\index{atom chip}Atom chips \cite{reichel_vuletic_book_11} facilitate controlled experiments with  \index{interaction(s)!weak}\index{weakly-interacting|see{interactions, weak}}weakly-interacting, effectively \index{one-dimensional}one-dimensional Bose gases \cite{schumm_hofferberth_05,esteve_trebbia_06,trebbia_esteve_06,hofferberth_lesanovsky_06,jo_choi_07,jo_shin_07,hofferberth_lesanovsky_07,vanamerongen_vanes_08,hofferberth_lesanovsky_08,baumgartner_sewell_10,armijo_jacqmin_10,manz_bucker_10,armijo_jacqmin_11}, including {\it in situ} measurements, thus allowing for precision tests against theory. Importantly, \index{fluctuations!thermal}fluctuations play a key role over a wide temperature regime for such extremely elongated geometries, thus making the \index{Gross--Pitaevskii equation! stochastic}SGPE a prime candidate for modelling such systems \cite{cockburn_gallucci_11}.
Figure~\ref{fig:q1d} shows a comparison between the \index{Gross--Pitaevskii equation! stochastic}SGPE and several, independent
sets of quasi-1d experimental data. The agreement is excellent when comparing both density profiles 
[Fig.\ref{fig:q1d}(a)--(b)] and \index{fluctuations!density}density \index{fluctuations!thermal}fluctuations [Fig.\ref{fig:q1d}(d)];
the latter requires the binning of {\it raw} \index{Gross--Pitaevskii equation! stochastic}SGPE data [see Fig.\ref{fig:q1d}(c)] to 
bins of width set by the resolution in a given experiment, 
a key step required to achieve a consistent analysis
of \index{fluctuations!thermal}fluctuations.
Experimentally, a system is considered as \index{one-dimensional}one-dimensional if 
$\mu,k_{\mathrm{B}}T\ll\hbar\omega_{\perp}$. If the first requirement is not satisfied, 
the replacement $g_{\rm 1d}|\psi|^2\rightarrow\hbar\omega_{\perp}[\sqrt{1+4a|\psi|^2}-1]$
\cite{fuchs_leyronas_03,gerbier_04,mateo_delgado_07} can account for the transverse swelling of the gas; 
in addition, if $k_{\mathrm{B}}T\not\ll\hbar\omega_{\perp}$, then we should also account for atoms in
transverse excited modes which contribute a density 
$n_{\perp}(x)=({1}/{\lambda_{\rm dB}})\sum_{j=1}^{\infty} (j+1) 
  {\rm g}_{1/2}(e^{[\mu-V_{\rm ext}(x)-j\hbar\omega_{\perp}]/k_{\mathrm{B}}T})$
\cite{vanamerongen_vanes_08} (${\rm g}_{1/2}(\bullet)$: polylogarithm of order $1/2$).
These two amendments yield a quasi-1d SGPE \cite{cockburn_gallucci_11}, which is \index{cutoff}cutoff independent (as both below and above \index{cutoff}cutoff physics is included in the model in an approximate, but {\it self-consistent} manner), and thus accurately models experiments in the crossover from one to three dimensions \cite{cockburn_gallucci_11,gallucci_cockburn_12}.

\subsection{\index{soliton!dark}Dark Soliton Dynamics in a \index{quasi-condensate}Quasi-Condensate \label{cockburn_application2}}

\begin{figure}[t]
  \begin{center}
    \includegraphics[angle=0,scale=0.25,clip]{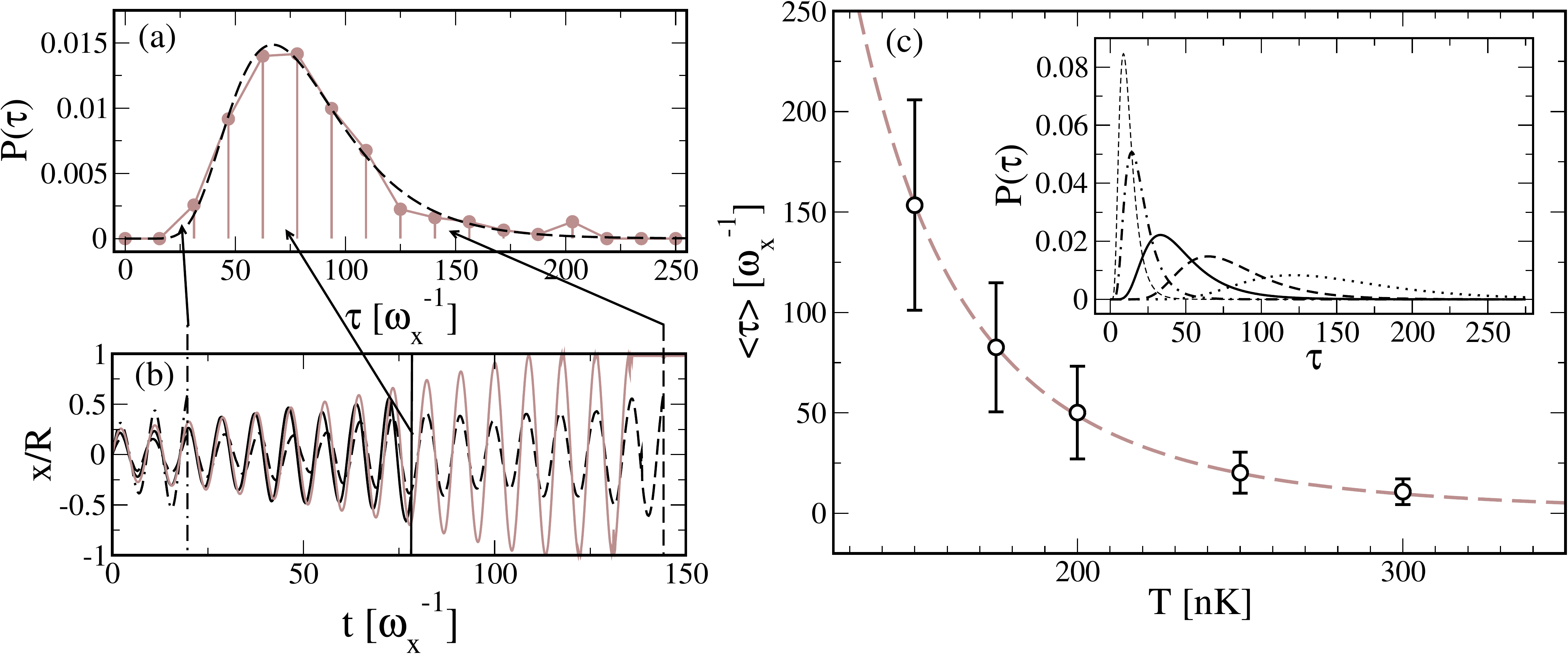}
    \caption[]{(a) Histogram of \index{soliton!dark}soliton decay times and (b) example \index{Gross--Pitaevskii equation! stochastic}SGPE \index{soliton!dark}soliton trajectories at $T=175$nK, from the indicated bins, shown by vertical dot-dashed, solid and dashed black lines; the solid grey trajectory corresponds to the \index{Gross--Pitaevskii equation! dissipative}DGPE result.
 (c) Average decay times extracted from \index{soliton!dark}soliton decay time histograms at several temperatures, the fits to which are shown in the inset.
Subplots adapted with permission from
S.P. Cockburn {\em et al.}, {\em Matter-Wave Dark Solitons: Stochastic versus Analytical Results},
Phys. Rev. Lett. {\bf 104}, 174101 (2010) \cite{cockburn_nistazakis_10}
and
S.P. Cockburn {\em et al.}, {\em Fluctuating and dissipative dynamics of dark solitons in quasi-condensates}, Phys. Rev. A {\bf 84}, 043640 (2011) \cite{cockburn_nistazakis_11}.
Copyright (2010) and (2011) by the American Physical Society.
}
	\label{fig:soli_fig}
  \end{center}	
\end{figure}

As the equilibrium state of the \index{Gross--Pitaevskii equation! stochastic}SGPE agrees well with both 
experiment \cite{cockburn_gallucci_11}
and alternative theories in suitable limits \cite{cockburn_negretti_11},
it constitutes an ideal {\it ab initio} approach for 
finite temperature  Bose
gas dynamics 
\cite{proukakis_schmiedmayer_06,cockburn_nistazakis_10,cockburn_thesis_10,cockburn_nistazakis_11} 
(see also \cite{proukakis_03,bradley_gardiner_08,weiler_neely_08,damski_zurek_10,rooney_bradley_10}).
This is particularly true 
for perturbations which do not push the \index{thermal!cloud}thermal cloud
\index{equilibrium! far from}far from equilibrium, making the dynamics of a \index{soliton!dark}dark soliton a perfect candidate.
Including \index{fluctuations!thermal}fluctuations in the background field leads to `shot-to-shot' variations in \index{soliton!dark}soliton behaviour, as evident from 
indicative trajectories shown in Fig.~\ref{fig:soli_fig}(b).
The corresponding histogram of decay times over the \index{ensemble! of realisations}ensemble of realisations 
(Fig.~\ref{fig:soli_fig}(a)) is
well-fitted by a lognormal distribution,
displaying an extended tail at 
long times, indicative of very long-lived \index{soliton!dark}solitons, relative to 
the average \index{soliton!dark}soliton decay time \cite{cockburn_nistazakis_10,cockburn_thesis_10,cockburn_nistazakis_11}, consistent with experiments \cite{becker_stellmer_08}.
These distributions are shown for a
range of temperatures in Fig.~\ref{fig:soli_fig}(c) (inset),
with the average times found to vary with temperature
as $\langle\tau\rangle\sim T^{-4}$ (main plot).
The distribution of \index{soliton!dark}soliton decay times obtained via the \index{Gross--Pitaevskii equation! stochastic}SGPE, 
relative to the single mean field \index{Gross--Pitaevskii equation! dissipative}DGPE result, indicates that consideration of many realisations of 
the \index{stochastic!wavefunction|see{stochastic, field}}\index{stochastic!field}stochastic wavefunction allows one to construct a representation 
of the full probability distribution for the gas;
moreover, our analysis
highlights once more the important feature 
that useful information is also retained within 
{\it individual} \index{Gross--Pitaevskii equation! stochastic}SGPE runs, which thus bear strong similarities to single-shot experimental
realisations\index{stochastic!trajectory interpretation} \cite{duine_stoof_01,weiler_neely_08,cockburn_nistazakis_10,cockburn_thesis_10,cockburn_gallucci_11,cockburn_nistazakis_11}.

\section*{Acknowledgments}

Nick Proukakis is indebted to Henk Stoof for stimulating this line of research and for an extended collaboration, and to Keith Burnett, Matt Davis, Allan Griffin, Carsten Henkel and Eugene Zaremba for extended discussions.
We also thank the EPSRC for funding, M.~Bijlsma, R.~Duine, D.~Frantzeskakis, D.~Gallucci, T.~Horikis, P.~Kevrekidis, A.~Negretti, H.~Nistazakis, T.~Wright for discussions and I.~Bouchoule, K.~van Druten and A.~van Amerongen for experimental data. 

\bibliographystyle{apsrev4-1}
\bibliography{cockburn_proukakis.bib}

\end{document}